\documentclass[fleqn,usenatbib]{mnras}

\usepackage{newtxtext,newtxmath}

\usepackage[T1]{fontenc}

\usepackage{listings}

\DeclareRobustCommand{\VAN}[3]{#2}
\let\VANthebibliography\thebibliography
\def\thebibliography{\DeclareRobustCommand{\VAN}[3]{##3}\VANthebibliography}

\usepackage{graphicx}	
\usepackage{amsmath}	
\usepackage{subfigure}

\title[Constraining NS Crust Shallow Heating via the Superburst Ignition Depth]{Constraining Accreted Neutron Star Crust Shallow Heating with the Inferred Depth of Carbon Ignition in X-ray Superbursts}

\author[Z. Meisel]{
Zach Meisel\thanks{E-mail: zachary.meisel@us.af.mil}
\\
Department of Engineering Physics, Air Force Institute of Technology, Wright-Patterson Air Force Base, Ohio 4533, USA\\
}

\date{Accepted XXX. Received YYY; in original form ZZZ}

\pubyear{2024}

\begin{document}
\label{firstpage}
\pagerange{\pageref{firstpage}--\pageref{lastpage}}
\maketitle

\begin{abstract}
Evidence has accumulated for an as-yet unaccounted for source of heat located at shallow depths within the accreted neutron star crust. However, the nature of this heat source is unknown. I demonstrate that the inferred depth of carbon ignition in X-ray superbursts can be used as an additional constraint for the magnitude and depth of shallow heating. The inferred shallow heating properties are relatively insensitive to the assumed crust composition and carbon fusion reaction rate. For low accretion rates, the results are weakly dependent on the duration of the accretion outburst, so long as accretion has ensued for enough time to replace the ocean down to the superburst ignition depth. For accretion rates at the Eddington rate, results show a stronger dependence on the outburst duration. Consistent with earlier work, it is shown that urca cooling does not impact the calculated superburst ignition depth unless there is some proximity in depth between the heating and cooling sources.
\end{abstract}




\section{Introduction}
Accreting neutron stars are unique probes of matter at high density and relatively low temperature, as well as extreme neutron-proton asymmetries~\citep{Fuku11,Meis18}. A number of observables provide unique insight into the nature of these ultradense objects, including X-ray bursts, X-ray superbursts, and crust cooling after accretion outbursts~\citep{Wijn17,intZ17,Gall21}. Model-observation comparisons are beginning to provide a handle on bulk properties of the underlying neutron star, such as the mass and radius, as well as evidence for exotic processes and phases of matter deep in the neutron star crust and even into the core~\citep[e.g][]{Brow09,Page13,Deib17,Cumm17,Brow18,Meis19,Good19}. While there have been several successes in modeling the aforementioned observed phenomena for various accreting neutron star sources, many require the addition of a heat source in the neutron star outer layers. Shallow heating has been used to explain the characteristic break in the crust cooling light curve~\citep{Dege14,Turl15,Merr16,Pari17,Page22}, the existence of short-waiting time bursts in X-ray bursting systems~\citep{Keek17}, and is likely necessary to resolve the discrepancy between modeled and inferred superburst ignition depths~\citep{Coop09}.

The physical mechanism for the shallow heat source is not known, which is one of the major outstanding problems in accreting neutron star physics~\citep{Scha22}. Nuclear reactions are known to be an important crustal heat source~\citep{Gupt07,Gupt08,Stei12}; however, both the depth and magnitude of this heat appear to be inconsistent with observational constraints for shallow heating~\citep{Brow09,Deib15,Fant18,Cham20}. Other suggested heat sources are related to compositionally driven convection in the accreted neutron star ocean and transfer of energy from the accretion flow to deeper depths via gravity waves~\citep{Inog10,Medi15,Deib15}. Determining if any of these explanations, or some combination thereof, ultimately suffice will require concerted model-observation comparison efforts.

Crust cooling model-observation comparisons likely provide the most stringent constraints on shallow heating, as the queiscent cooling light curve provides a tomographic picture of the accreted crust~\citep{Page13}. However, these model calculations have a large number of poorly-constrained parameters and therefore many degenerate solutions as to the strength and depth of shallow heating. As such, complementary constraints on the properties of shallow heating in accreting neutron star crusts are desirable. The inferred depth of carbon ignition for X-ray superbursts provides such an opportunity.

X-ray superbursts are thought to be energetic explosions ignited by carbon fusion in the accreted neutron star ocean and primarily powered by the photodisintegration of heavy nuclei remaining from earlier surface burning~\citep{Taam78,Cumm01,Stro02,Corn03,Scha03}. The ignition column depth can be roughly inferred based on the typical recurrence time $\Delta t_{\rm rec}\sim1$~yr and accretion rate $\dot{M}\sim5\times10^{-9}$~M$_{\odot}/$yr for superbursting systems~\citep{intZ03,Gall20} as $y_{\rm ign}=\dot{M}\Delta t_{\rm rec}/\left(4\pi R_{\rm NS}^{2}\right)\sim5\times10^{11}$~g\,cm$^{-2}$~\citep{Meis18}, assuming a neutron star radius $R_{\rm NS}\sim12$~km~\citep{Rile21}. A more rigorous analysis based on fitting the observed superburst light curve with cooling models results in the ignition column depth inferred from observations, $y_{\rm ign,obs}=0.5-3\times10^{12}$~g\,cm$^{-2}$~\citep{Cumm06}. This range for $y_{\rm ign,obs}$ is somewhat sensitive to the neutron star envelope temperature profile that is assumed prior to the superburst~\citep{Keek15}, but this correction is not considered here.

These constraints on the superburst $y_{\rm ign,obs}$ can be confronted with results from model calculations of carbon ignition in the accreted neutron star ocean. As is described in more detail in the following sections, ignition curves based on adopted heating and cooling rates can be paired with models of the accreted neutron star thermal structure in order to calculate $y_{\rm ign}$. Comparisons of the calculated and inferred $y_{\rm ign}$ can place constraints on the accreted neutron star thermal structure and therefore on the magnitude and depth of shallow heating.

In this work, I perform comparisons of calculated and inferred $y_{\rm ign}$ in order to place constraints on the properties of the shallow heat source thought to be present in accreted neutron star crusts. In Section~\ref{sec:calculations}, I describe the calculations of the carbon ignition curves and crust thermal profiles, as well as the superburst ignition depth. Section~\ref{sec:results} contains the calculation results. In Section~\ref{sec:discussion}, the constraints on the shallow heat source depth and magnitude are discussed, followed by a discussion of the nuclear physics uncertainties potentially impacting the results, as well as a discussion of incorporating the technique presented here into future multi-observable model-observation comparisons. Section~\ref{sec:conclusions} contains a summary.

\section{Calculations}
\label{sec:calculations}
This work follows the $y_{\rm ign}$ calculation approach of \citet{Deib16}, who closely followed the method presented by~\citet{Pote12}. For a chosen $^{12}$C+$^{12}$C fusion rate and ocean thermal conductivity, the changes in the nuclear energy generation rate and in the cooling rate with a change in temperature are calculated and, at each column depth, it is determined what temperature is required for the heating derivative to exceed the cooling derivative. For an adopted set of astrophysical conditions and crust microphysics, the temperature as a function of depth is determined by numerically solving the general relativistic heat diffusion equation, in this case using the code {\tt dStar}~\citep{Brow15}. The point at which an ignition curve intersects a thermal profile is the superburst ignition depth for that set of astrophysical conditions and microphysics. Each of these steps is described in more detail in the following subsections.

\subsection{Carbon Ignition Curves}
\label{ssec:carbonignition}
Nuclear energy generation at the ignition of a superburst is set by the $^{12}$C+$^{12}$C fusion rate. At temperatures relevant for the accreted neutron star envelope, this nuclear reaction rate is based on nuclear theory calculations and is uncertain by several orders of magnitude~\citep{Beck20,Tang22,Alio22}. Modern theoretical $^{12}$C+$^{12}$C rates include results from barrier penetration calculations using the Sao Paulo potential~\citep{Yako10}, coupled-channel calculations performed using the M3Y+repulsion double-folding potential~\citep{Esbe11}, empirical extrapolations based on the hindrance model~\citep{Jian18}, experimentally derived results based on the trojan horse method (THM)~\citep{Tumi18}, THM results adopting a Coulomb renormalization~\citep{Mukh19}, and a microscopic model with molecular resonances~\citep{Tani21}. Each of these are used in the present work.

Theoretical results for $^{12}$C+$^{12}$C fusion are typically presented as a modified astrophysical $S$-factor, $S^{*}$, where the $S$-factor is $S(E)=S^{*}(E)\exp(-0.46E)$, with $E$ as the center-of-mass energy of the reaction. For nuclear reactions involving the fusion of two charged particles, the $S$-factor is related to the directly measured (or calculated) cross section $\sigma(E)$ by $\sigma(E)=S(E)\exp(-2\pi\eta)/E$, where $\eta$ is defined below. Following~\citet{Pote12}, the thermonuclear fusion rate of nuclear species 1 and 2 per unit volume at temperature $T$ in an electron-degenerate environment characterized by electron chemical potential $\mu_{e}$ is
\begin{equation}
\begin{split}
\mathcal{R}_{12}(T,\mu_{e})=\frac{w_{12}c\sqrt{8}}{\sqrt{\pi \mu_{\rm red}m_{u}\left(k_{\rm B}T\right)^{3}}}n_{1}(\mu_{e})n_{2}(\mu_{e}){\rm INT},\\
{\rm INT}=\int_{0}^{\infty}S(E_{\rm s})\exp\left(-2\pi\eta\left(E_{\rm s}\right)-E/(k_{\rm B}T)\right)dE.
\label{eqn:reactionrate}
\end{split}
\end{equation}
\noindent Here, $c$ is the speed of light in vacuum, $k_{\rm B}$ is the Boltzmann constant, $m_{u}$ is the nucleon mass, $w_{12}=0.5$ for identical nuclear species and 1 otherwise, and $\mu_{\rm red}=(A_{1}A_{2})/(A_{1}+A_{2})$ is the reduced mass of species with nuclear mass numbers $A_{i}$. The species' number densities $n_{i}$ are related to $\mu_{e}$ via the mass-density $\rho$ by $n_{i}=(X_{i}/A_{i})\rho$, where, for an environment in which the pressure is dominated by degenerate electrons~\citep{Meis18},
\begin{equation}
\rho(\mu_{e})\approx\frac{7.2\times10^{6}}{Y_{e}}\left(\frac{\mu_{e}}{1\,{\rm MeV}}\right)^{3}\,{\rm g}\,{\rm cm}^{-3}.
\end{equation}
 The electron fraction $Y_{e}=\sum_{i}Z_{i}X_{i}/A_{i}$ is summed over all species at $\mu_{e}$, where each species has nuclear charge $Z_{i}$ and mass-fraction $X_{i}$. The Sommerfeld parameter is
$\eta(E)=\sqrt{(Z_{1}^{2}Z_{2}^{2}\alpha_{\rm fs}^{2}\mu_{\rm red}m_{u})/(2E)}$, where $\alpha_{\rm fs}$ is the fine-structure constant. In order to account for the enhancement of the fusion rate due to plasma screening, both $S(E)$ and $\eta(E)$ are evaluated at a shifted energy, $E_{\rm s}=E+H_{12}(0)$~\citep{Cler19}. The temperature-dependent energy shift $H_{12}(0)$ for a high-density environment can be approximated as $H_{12}(0)=k_{\rm B}Th_{12}^{0}$, where $h_{12}^{0}=f_{0}(\Gamma_{i})+f_{0}(\Gamma_{j})-f_{0}(\Gamma_{ij}^{\rm comp})$, assuming the linear mixing rule~\citep{Chug09}. The terms $f_{0}(\Gamma)$ are the Coulomb free energy per ion in a one component plasma using the analytic approximation~\citep{Pote00}

\begin{equation}
\begin{split}
f_{0}(\Gamma)=&a_{1}\left(\sqrt{\Gamma(a_{2}+\Gamma)}-a_{2}\ln\left(\sqrt{\Gamma/a_{2}}+\sqrt{1+\Gamma/a_{2}}\right)\right)\\
&+2a_{3}\left(\sqrt{\Gamma}-\arctan(\sqrt{\Gamma})\right)+b_{1}\left(\Gamma-b_{2}\ln(1+\Gamma/b_{2})\right)\\
&+(b_{3}/2)\ln(1+\Gamma^{2}/b_{4}),
\end{split}
\end{equation}
\noindent where $a_{1}=-0.907$, $a_{2}=0.62954$, $a_{3}=0.2771$, $b_{1}=0.00456$, $b_{2}=211.6$, $b_{3}=-10^{-4}$, and $b_{4}=0.00462$. The ion coupling radii are $\Gamma_{i}=\alpha_{\rm fs}\hbar cZ_{i}^{5/3}/(a_{e}k_{\rm B}T)$, where $\hbar$ is the reduced Planck constant and the electron sphere radius is $a_{e}=\left(3/(4\pi n_{e}(\mu_{e}))\right)^{1/3}$. The electron number density at $\mu_{e}$ is $n_{e}(\mu_{e})=(Y_{e}/m_{u})\rho(\mu_{e})$. For the compound nucleus resulting from the fusion of species $1+2$, $Z_{i}=Z_{1}+Z_{2}$. 

The local nuclear energy generation rate per unit mass is
\begin{equation}
\epsilon_{\rm nuc}=\mathcal{R}_{12}Q_{12}/\rho,
\label{eqn:enuc}
\end{equation}
where $Q_{12}$ is the energy release of a fusion event between species 1 and 2.

The present work deals with $^{12}$C+$^{12}$C fusion, so $w_{12}=0.5$, $Z_{1}=Z_{2}=6$, $A_{1}=A_{2}=12$, $Q_{12}$ is the absolute value of the atomic mass excess of the compound nucleus $^{24}{\rm Mg}$ ($|{\rm ME}(^{24}{\rm Mg})|=13.9336$~{\rm MeV}\citep{Wang21}), and the adopted $S^{*}$ come from the aforementioned nuclear theory calculations. The envelope is assumed to be comprised of only $^{12}{\rm C}$ and $^{56}{\rm Fe}$, with $X_{\rm C}=0.2$ and $X_{\rm Fe}=0.8$.

The local cooling rate per unit mass from thermal diffusion is \citep{Fuji81,Pote12}
\begin{equation}
\epsilon_{\rm cool}=\kappa_{\rm eff}\rho T/y^{2},
\label{eqn:ecool}
\end{equation}
where $\kappa_{\rm eff}=0.17\kappa$ is the effective thermal conductivity. The thermal conductivity $\kappa=\pi^{2}c^{2}k_{\rm B}^{2}Tn_{e}/(3\mu_{e}\nu_{\rm coll})$~\citep{Meis18}, where the collision frequency $\nu_{\rm coll}$ in a liquid ocean is determined by the electron-ion impurity using the linear mixing rule approximation~\citep{Brow04}: $\nu_{\rm coll}=4\alpha_{\rm fs}^{2}\mu_{e}\langle Z^{2}\Lambda_{\rm ei}\rangle/(3\pi\hbar\langle Z\rangle)$. Here, $\langle\rangle$ are mass-fraction-weighted averages of the composition at $\mu_{e}$ and $\Lambda_{\rm ei}=1$ is the Coulomb logarithm, where it is noted that a more accurate estimate of $\Lambda_{\rm ei}$ would consider the accreted neutron star envelope structure~\citep{Horo09,Roge16}. For an environment in which the pressure is dominated by degenerate electrons, the column depth $y$ is related to $\mu_{e}$ by~\citep{Meis18}
\begin{equation}
y\approx7.2\times10^{9}\left(\frac{\mu_{e}}{1\,{\rm MeV}}\right)^{4}\frac{2.44\times10^{14}\,{\rm cm}\,{\rm s}^{-2}}{g}\,{\rm g}\,{\rm cm}^{-2}.
\label{eqn:coldepth}
\end{equation}
The local gravitational acceleration is $g=(GM_{\rm NS}/R_{\rm NS}^{2})(1+z)$, where $G$ is the gravitational constant, $M_{\rm NS}$ is the neutron star mass, and the surface gravitational redshift is $(1+z)=1/\sqrt{1-2GM_{\rm NS}/(R_{\rm NS}c^{2})}$.

\begin{figure}
	\includegraphics[width=\columnwidth]{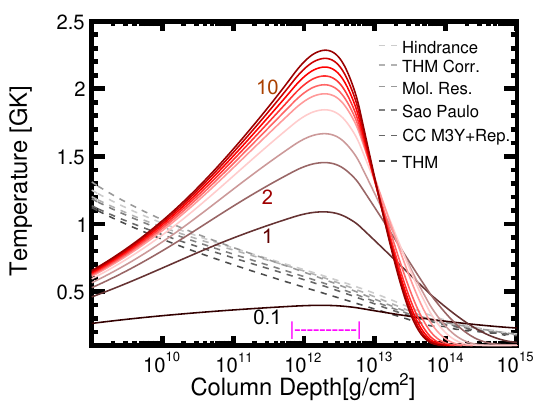}
    \caption{Carbon ignition curves (gray dashed-lines) for $^{12}$C+$^{12}$C fusion rates used in this work: Hindrance~\citep{Jian18}, THM Corr.~\citep{Mukh19}, Mol. Res.~\citep{Tani21}, Sao Paulo~\citep{Yako10},  CC M3Y+Rep~\citep{Esbe11}, and THM~\citep{Tumi18}. The order of rates in the legend is the order (from high-to-low $T$) of the ignition curves at $y=10^{12}$~g\,cm$^{-2}$. Thermal profiles (red solid lines) calculated with {\tt dStar} for the maximum urca cooling scenario are also shown, where $\dot{M}=1.75\times10^{-8}$~$M_{\odot}$\,yr$^{-1}$, $Q_{\rm imp}=4$, and $\Delta t=1643.6$~d. Shallow heating was deposited in the range indicated by the pink $|---|$, with $Q_{\rm sh}=0.1$~MeV\,$u^{-1}$ and $1-10$~MeV\,$u^{-1}$ in steps of 1~MeV, as indicated by the labels next to the profiles.}
    \label{fig:ExampleProfiles}
\end{figure}

Thermal instability for superburst ignition is achieved when the change in the nuclear heating rate outpaces the change in the cooling rate with a change in temperature~\citep{Fush87}:
\begin{equation}
\frac{\partial\epsilon_{\rm nuc}}{\partial T}>\frac{\partial\epsilon_{\rm cool}}{\partial T}.
\label{eqn:ignition}
\end{equation}
The carbon ignition curves shown in Figure~\ref{fig:ExampleProfiles} were calculated by identifying the minimum $T$ needed to satisfy this inequality at each $y$ located within part of the neutron star ocean and outer crust.

\subsection{Crust Thermal Profiles}
\label{ssec:thermalprofiles}
Thermal profiles (i.e. temperature as a function of depth) for the accreted neutron star crust were calculated for a large number of somewhat arbitrarily chosen but astrophysically relevant conditions using the open-source code {\tt dStar}~\citep{Brow15}. In {\tt dStar}, the thermal evolution of a neutron star undergoing (or having undergone) accretion is calculated by solving the general relativistic heat diffusion equation using the {\tt MESA}~\citep{Paxt11,Paxt13,Paxt15} numerical libraries with the microphysics described by \citet{Brow09}. The input file for a {\tt dStar} calculation is known as an {\it inlist}, as shown in Appendix~\ref{sec:appendixA}, which can be used to specify a number of astrophysical parameters, microphysics models, and numerical controls. Key input quantities are described in this subsection. In addition, Tables~\ref{tab:inputs} and~\ref{tab:urca} list the input parameters that were varied between calculations.

Accretion drives the neutron star outer layers out of thermal equilibrium with the core, where heat is deposited into the crust via nuclear reactions that are driven by accretion, which can also lead to neutrino cooling~\citep{Lau18,Scha22b}. The thermal profile, and thereby the temperature at each radial coordinate $r$  (assuming spherical symmetry), over time $t$ is determined via the heat diffusion equation~\citep{Page13}:
\begin{equation}
C_{V}\frac{\partial T}{\partial t}=\kappa\frac{\partial^{2}T}{\partial r^{2}}+\frac{1}{r^{2}}\frac{\partial(r^{2}\kappa)}{\partial r}\frac{\partial T}{\partial r}
\label{eqn:diffusion} + Q_{\rm heat} - Q_{\rm cool},
\end{equation}
where the specific heat $C_{V}$ (described in detail by~\citet{Brow09}), $\kappa$, nuclear heating rate $Q_{\rm heat}$, and neutrino cooling rate $Q_{\rm cool}$ are each depth dependent. The neutron star core can be approximated as an infinite heat sink, though this is not quite the case~\citep{Cumm17,Brow18}, fixing the temperature at the crust-core boundary to the core temperature $T_{\rm core}$. In this work,  $T_{\rm core}=10^{8}$~K based on typical constraints for several crust-cooling systems~\citep{Page13,Dege14,Deib15,Lali19}.

In the crust, $\kappa$ is related to the variance in the nuclear charge of the composition, as described in Section~\ref{ssec:carbonignition}; however, the composition is generally not the same as in the ocean. It is customary to describe the charge variance in the accreted neutron star crust by the impurity parameter~\citep{Itoh93,Brow09}:
\begin{equation}
Q_{\rm imp}=\frac{\sum_{i}n_{i}\left(Z_{i}-\langle Z\rangle\right)^{2}}{\sum_{i}n_{i}}.
\label{eqn:qimp}
\end{equation}
In reality, $Q_{\rm imp}$ evolves over depth due to nuclear reactions and due to changing surface burning over time, e.g. because of changing $\dot{M}$. However, both the nuclear physics of crust reactions and the surface burning history of a given accreting neutron star system have significant uncertainties. It is therefore more common to employ a single approximate $Q_{\rm imp}$ in model-observation comparisons. In this work, $Q_{\rm imp}=4$ and 40 are adopted, where the former has successfuly explained crust-cooling observations of the superbursting system KS 1731-26~\citep{Lali19} and the latter is the largest $Q_{\rm imp}$ yet used to reproduce any observed crust-cooling light curve~\citep{Dege14}.

Here, deep crustal heating is approximated by depositing 1.5~MeV per accreted nucleon (MeV\,$u^{-1}$) across $y=5\times10^{15}-2\times10^{17}$~g\,cm$^{-2}$ and $e^{-}$-capture heating is approximated by depositing 0.3~MeV\,$u^{-1}$ across $y=5\times10^{12}-2\times10^{15}$~g\,cm$^{-2}$, consistent with recent estimates~\citep{Gupt08,Haen08}. Absent a physical model, shallow heating of strength $Q_{\rm sh}$ is deposited uniformly about a column depth $y_{\rm sh}$ within the range $y_{\rm sh}/3$ to $3y_{\rm sh}$, following~\citet{Deib15}. Each of the shallow heating magnitudes and depths listed in Table~\ref{tab:inputs} was employed in combination with each of the other input parameter options. The range of  $Q_{\rm sh}$ adopted was 0.1-10~MeV\,$u^{-1}$ in steps of 0.1, where the lower-bound is equivalent to using a higher-end estimate for $e^{-}$-capture heating and the upper estimate is the maximum $Q_{\rm sh}$ thus far inferred from accreting neutron star model-observation comparisons~\citep{Deib15}. Rather than selecting the shallow heating depth in terms of $y_{\rm sh}$, this depth was selected on a grid of pressure $P_{\rm sh}$ over the range $\log(P_{\rm sh})=24-29$, in cm-g-s units, in steps of 0.05. This is intended to encompass the range of $y_{\rm sh}$ found to be plausible by crust cooling model-observation comparisons~\citep{Deib15,Merr16,Pari18,Oote19,Page22}. For this depth range, where the pressure is primarily due to electron degeneracy, the pressure $P=\mu_{e}^{4}/(12\pi^{2}\hbar^{3}c^{3})$~\citep{Meis18}.

The total amount of heat deposited into the accreted neutron star outer layers during an accretion outburst depends on the duration $\Delta t$ and average $\dot{M}$ of the accretion outburst. The two values for $\dot{M}$ used in this work are approximately 10\% and 100\%, respectively, of the Eddington accretion rate for a standard neutron star accreting hydrogen-rich fuel~\citep{Scha99}. The smaller accretion rate is in the range typically inferred for superbursting systems~\citep{intZ17} and the larger is roughly the accretion rate at which stable burning begins~\citep{Gall21}. Here, {\tt dStar} thermal profiles were recorded for each calculation at $\Delta t$ of 1643.6~d and 4565~d. The former is the $\Delta t$ required to replace the envelope down to $y=10^{12}$~g\,cm$^{-2}$ at 10\% Eddington accretion rate, while the latter is the duration of the accretion outburst observed for KS 1731-26 prior to going into quiescence in 2001~\citep{Merr16}. Neither of these $\Delta t$ are sufficient to reach a steady-state temperature profile, but the longer of the two is close to achieving that state~\citep{Page13}. 

Neutrinos can be emitted from spherical shells in the crust via $e^{-}$-capture/$\beta^{-}$-decay cycling, known as urca cooling~\citep{Scha14}. The neutrino luminosity $L_{\nu}$ associated with $e^{-}$-capture parent species ($Z_{i}$,$A_{i}$) with mass fraction $X_{i}$ is~\citep{Tsur70,Deib16}:
\begin{equation}
L_{\nu,i} \approx L_{34}\times10^{34}{\rm{erg\,s}}{}^{-1}X_{i}T_{9}^{5}\left(\frac{g_{14}}{2}\right)^{-1}R_{10}^{2} \ ,
\label{eqn:Lnu}
\end{equation}
where $T_{9}$ is the temperature of the urca shell in units of $10^{9} \, \mathrm{K}$, $R_{10}\equiv R_{i}/(10~\rm{km})$, $R_{i}\approx R_{\rm NS}$ is the radius of the urca shell from the neutron star center, and $g_{14}\equiv g/(10^{14}~\rm{cm}\,\rm{s}^{-2})$. $L_{34}(Z_{i},A_{i})$ is the intrinsic cooling strength:
\begin{equation}
L_{34}=0.87\left(\frac{10^{6}~{\rm{s}}}{ft}\right)\left(\frac{56}{A_{i}}\right)\left(\frac{|Q_{\rm{EC}}|}{4~{\rm{MeV}}}\right)^{5}\left(\frac{\langle F\rangle^{*}}{0.5}\right). 
\label{eqn:L34}
\end{equation}
The energy-cost for $e^{-}$-capture is the $e^{-}$-capture $Q$-value $Q_{\rm EC}={\rm ME}(Z_{i},A_{i})-{\rm ME}(Z_{i}-1,A_{i})$, where the atomic mass excesses ME are corrected by a Coulomb lattice energy $+C_{\ell}Z^{5/3}Q_{\rm EC,0}$, $Q_{\rm EC,0}$ is the $Q$-value without the lattice correction, and $C_{\ell}\approx3.407\times10^{-3}$~\citep{Roca08}. The factor $\langle F\rangle^{*}\equiv\langle F\rangle^{+}\langle
F\rangle^{-}/(\langle F\rangle^{+}+\langle F\rangle^{-})$, where the Coulomb factor $\langle F\rangle^{\pm}\approx2\pi\alpha_{\rm fs}
Z_{i}/|1-\exp(\mp2\pi\alpha_{\rm fs} Z_{i})|$. The comparative half-life of the weak transition $ft$ is the average for the $\beta^{-}$-decay and $e^{-}$-capture reactions in the urca cycle, $ft=(ft_{\beta}+ft_{\rm EC})/2$, where the two are related by the spin $J$ degeneracy of the initial states $ft_{\beta}/(2J_{\beta}+1)=ft_{\rm EC}/(2J_{\rm EC}+1)$~\cite{Paxt16}. In principle, Equation~\ref{eqn:L34} can be modified by the thermal population of nuclear excited states~\citep{Misc21}, but they are ignored in the present work.

For simplicity, the present work is limited to investigating the impact of urca cooling from $^{55}{\rm Sc}-^{55}$Ca, as this pair is thought to have more than an order of magnitude larger $L_{\nu}$ than the next most significant urca pair~\citep{Deib16,Meis17} when adopting superburst ashes, where $X(A=55)=0.018$~\citep{Scha14,Keek12}. Four sets of urca cooling conditions were investigated, summarized in Table~\ref{tab:urca}. For the no-cooling scenario, corresponding to an absence of $^{55}{\rm Sc}$ at the approprate depth ($\mu_{e}\approx Q_{\rm EC})$, $XL_{34}=0$. The nominal cooling scenario corresponds to using the current best estimates for $Q_{\rm EC,0}$ and $ft$, while the maximum and minimum cooling scenarios correspond to upper and lower limits calculated based on the uncertainties for these parameters. For consistency, the pressure at which the urca shell is located $P_{\rm urca}$ is modified corresponding to $Q_{\rm EC,0}$. In this work, $|Q_{\rm EC,0}|=12.192$~MeV is calculated using ME from \citet{Mich18,Leis21} and the associated uncertainty $\delta Q_{\rm EC,0}=0.172$~MeV is calculated using their one-standard-deviation uncertainties. I use $ft=5.9$ and associated uncertainty $\delta ft=2$ based on the systematics of \citet{Sing98}, as experimental constraints do not yet exist for this weak transition.

To summarize the presentation of the thermal profile calculations, $161\,600$ {\tt dStar} calculations were performed, using all combinations of the input parameters and urca cooling conditions detailed in Tables~\ref{tab:inputs} and \ref{tab:urca}, respectively. Example thermal profiles are shown in Figure~\ref{fig:ExampleProfiles}.

\begin{table}
	\centering
	\caption{{\tt dStar} input parameters which were varied for the large grid of calculations performed in this work, where ``cgs" indicates in cm-g-s units. Note that $\dot{M}$ and $Q_{\rm imp}$ each only had two settings. Urca cooling settings are described in Table~\ref{tab:urca}.}
	\label{tab:inputs}
	\begin{tabular}{lccc} 
		\hline
		Parameter & Lower Bound & Upper Bound & Step Size\\
		\hline
		$\dot{M}$ [$M_{\odot}$\,yr$^{-1}$] & 1.75$\times10^{-9}$ & 1.75$\times10^{-8}$ & -\\
		$Q_{\rm sh}$~[MeV\,$u^{-1}$] & 0.1 & 10 & 0.1\\
		$\log (P_{\rm sh}[\rm cgs])$ & 24 & 29 & 0.05\\
         $Q_{\rm imp}$ & 4 & 40 & -\\
		\hline
	\end{tabular}
\end{table}

\begin{table}
	\centering
	\caption{Urca cooling conditions used in {\tt dStar} calculations for this work, where ``cgs" indicates in cm-g-s units. All other varied inputs are described in Table~\ref{tab:inputs}.}
	\label{tab:urca}
	\begin{tabular}{lcc} 
		\hline
		Mode & $X(A)L_{34}$ & $\log (P_{\rm urca}[\rm cgs])$ \\
		\hline
		None & 0 &  -\\
		Minimum & 4.47$\times10^{-2}$ & 29.135\\
		Nominal & 4.80 & 29.159 \\
        Maximum & 5.15$\times10^{2}$ & 29.184 \\
		\hline
	\end{tabular}
\end{table}

\subsection{Superburst Ignition Depth}
\label{ssec:ignitiondepth}

For a single carbon ignition curve and thermal profile, $y_{\rm ign}$ is determined by numerically finding the intersection of the two. In a physical system, carbon will accumulate until a sufficient depth is reached for ignition, and as such the shallowest-$y$ intersection is the one of interest. The $y_{\rm ign}$ was determined for each combination of the six carbon ignition curves described in Section~\ref{ssec:carbonignition} and the $161\,600$ thermal profiles described in Section~\ref{ssec:thermalprofiles}.

\begin{figure*}
  \centering
  \subfigure[$Q_{\rm imp}=4$, $\dot{M}=1.75\times10^{-9}$~$M_{\odot}$\,yr$^{-1}$, Urca=None]{\label{fig:yignQ4M10}\includegraphics[width=1.35\columnwidth]{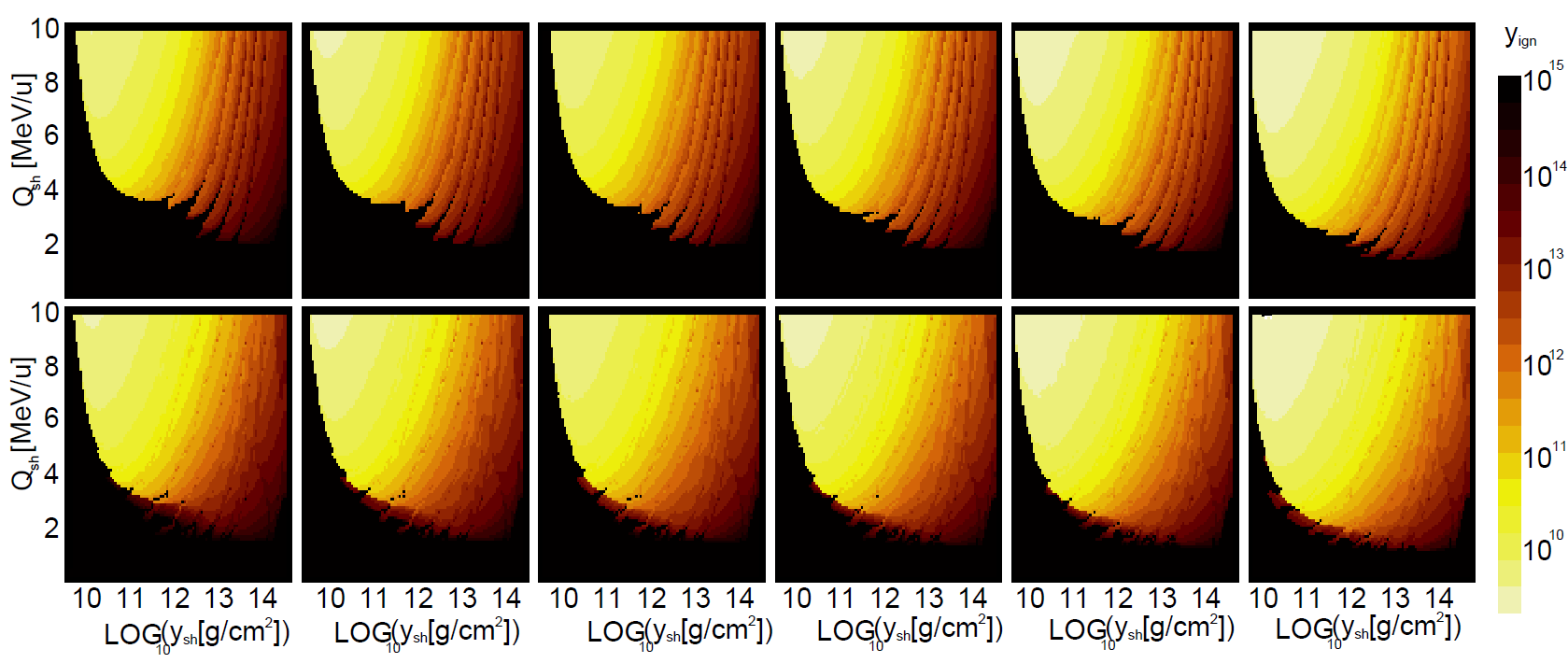}}
  \medskip
  \subfigure[$Q_{\rm imp}=40$, $\dot{M}=1.75\times10^{-9}$~$M_{\odot}$\,yr$^{-1}$, Urca=None]{\label{fig:yignQ40M10}\includegraphics[width=1.35\columnwidth]{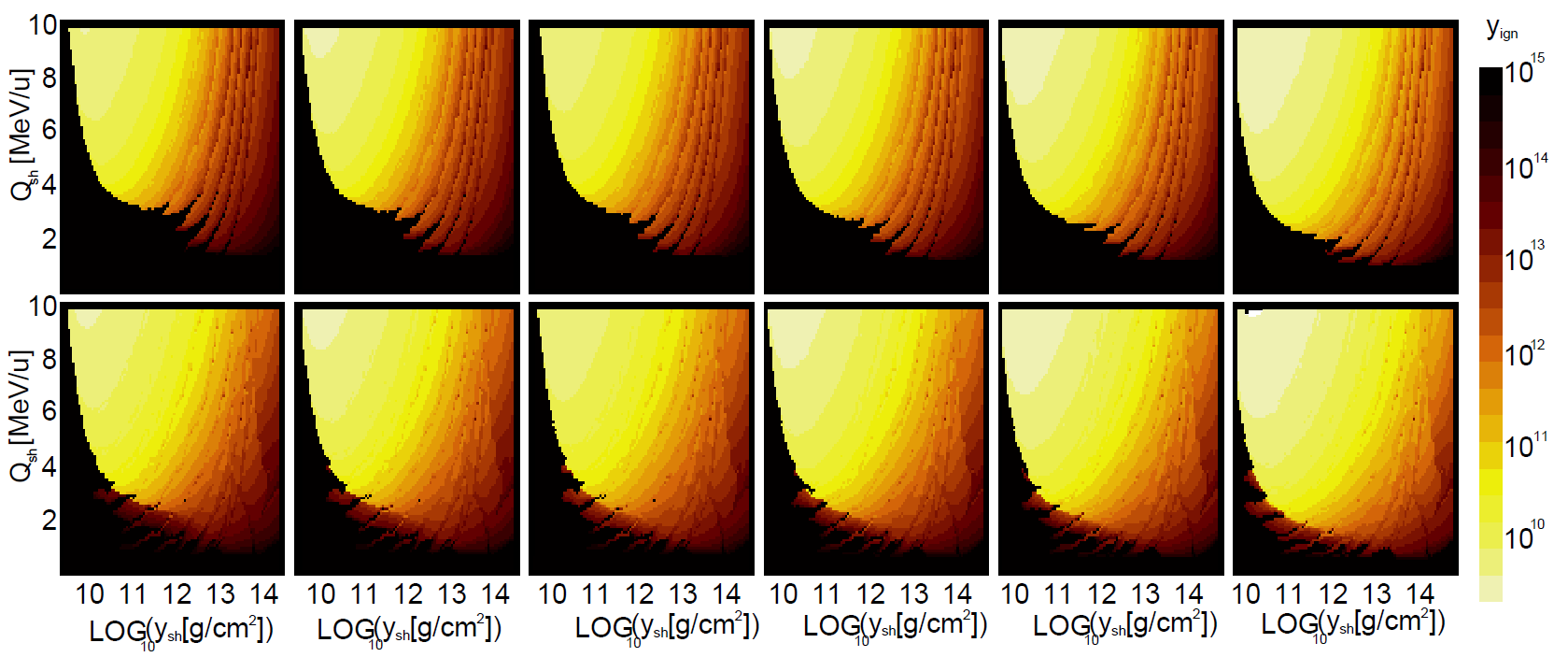}}
  \medskip
    \subfigure[$Q_{\rm imp}=4$, $\dot{M}=1.75\times10^{-8}$~$M_{\odot}$\,yr$^{-1}$, Urca=None]{\label{fig:yignQ4M100}\includegraphics[width=1.35\columnwidth]{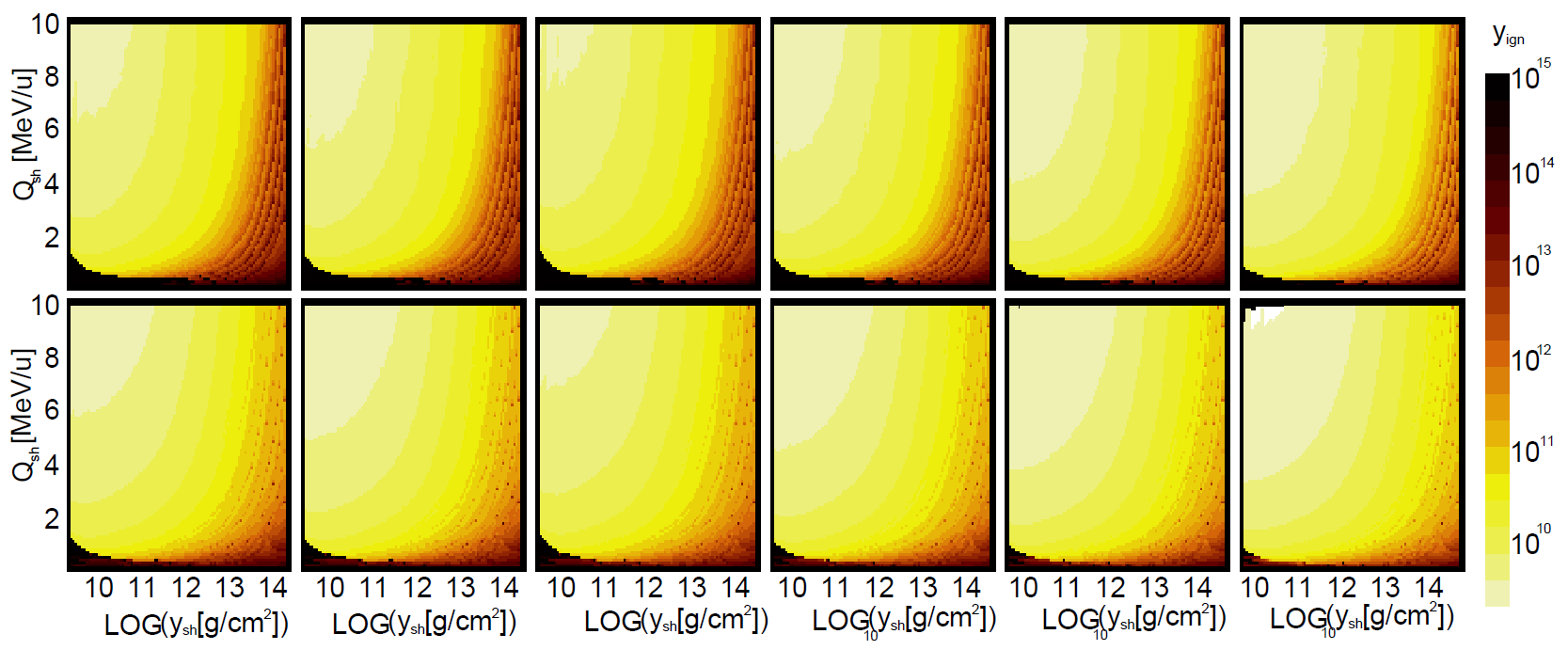}}
   
   \medskip
  \subfigure[$Q_{\rm imp}=40$, $\dot{M}=1.75\times10^{-8}$~$M_{\odot}$\,yr$^{-1}$, Urca=None]{\label{fig:yignQ40M100}\includegraphics[width=1.35\columnwidth]{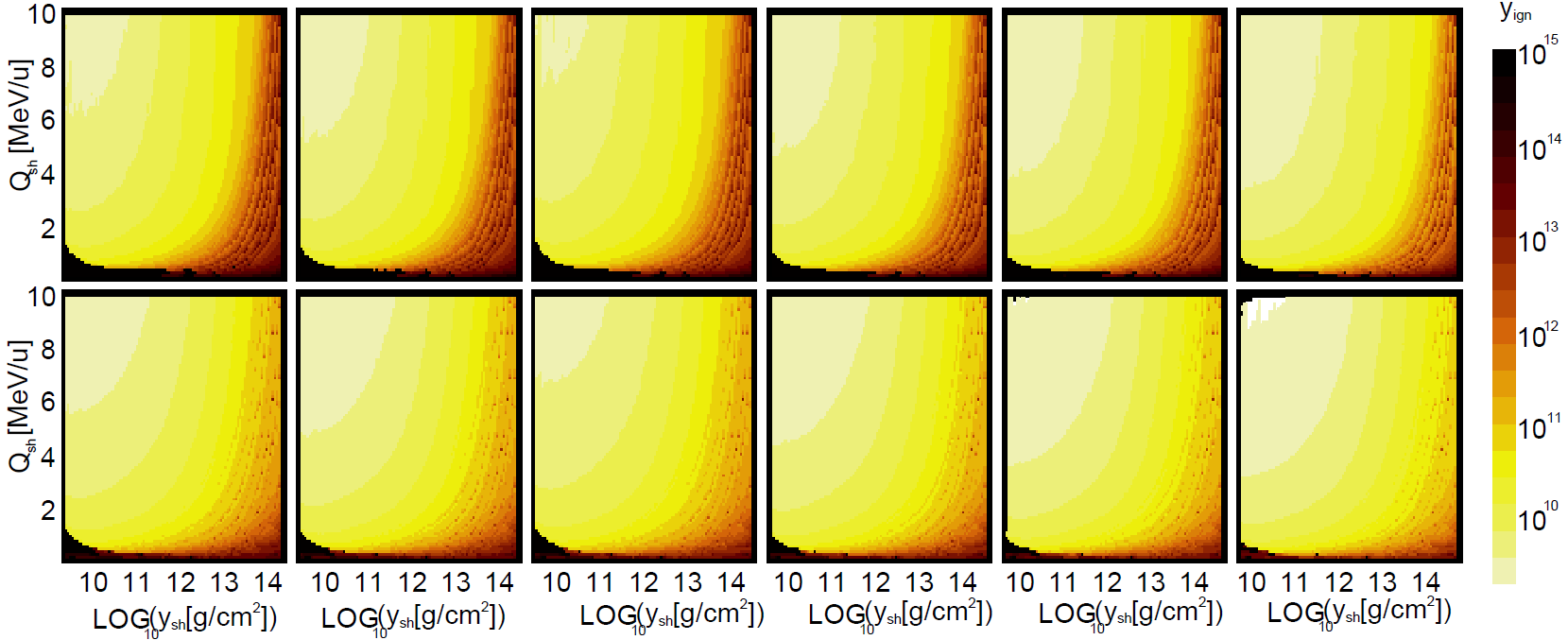}}
 \caption{Calculated $y_{\rm ign}$ in units of g\,cm$^{-2}$ (indicated by color, where black indicates that the conditions for $y_{\rm ign}$ were not satisfied) for the modeled range of $y_{\rm sh}$ and $Q_{\rm sh}$ using the indicated $Q_{\rm imp}$ and $\dot{M}$, without an urca cooling layer. Columns of figures correspond to a particular $^{12}{\rm C}+^{12}$C reaction rate, from left to right and using the labels of Figure~\ref{fig:ExampleProfiles}: Hindrance, THM Corr., Mol. Res., Sao Paulo, CC M3Y+Rep., THM. In each subfigure, the upper row corresponds to $\Delta t=1643.6$~d, while the lower row corresponds to $\Delta t=4565$~d.}
 \label{fig:yignmap}
\end{figure*}

\begin{figure*}
  \centering
  \subfigure[$Q_{\rm imp}=40$, $\dot{M}=1.75\times10^{-8}$~$M_{\odot}$\,yr$^{-1}$, Urca=None]{\label{fig:yignQ40M100}\includegraphics[width=2\columnwidth]{IgnitionDepths_GridQ40M100pctNoUrca.png}}
  \medskip
  \subfigure[$Q_{\rm imp}=40$, $\dot{M}=1.75\times10^{-8}$~$M_{\odot}$\,yr$^{-1}$, Urca=Maximum]{\label{fig:yignQ40M100}\includegraphics[width=2\columnwidth]{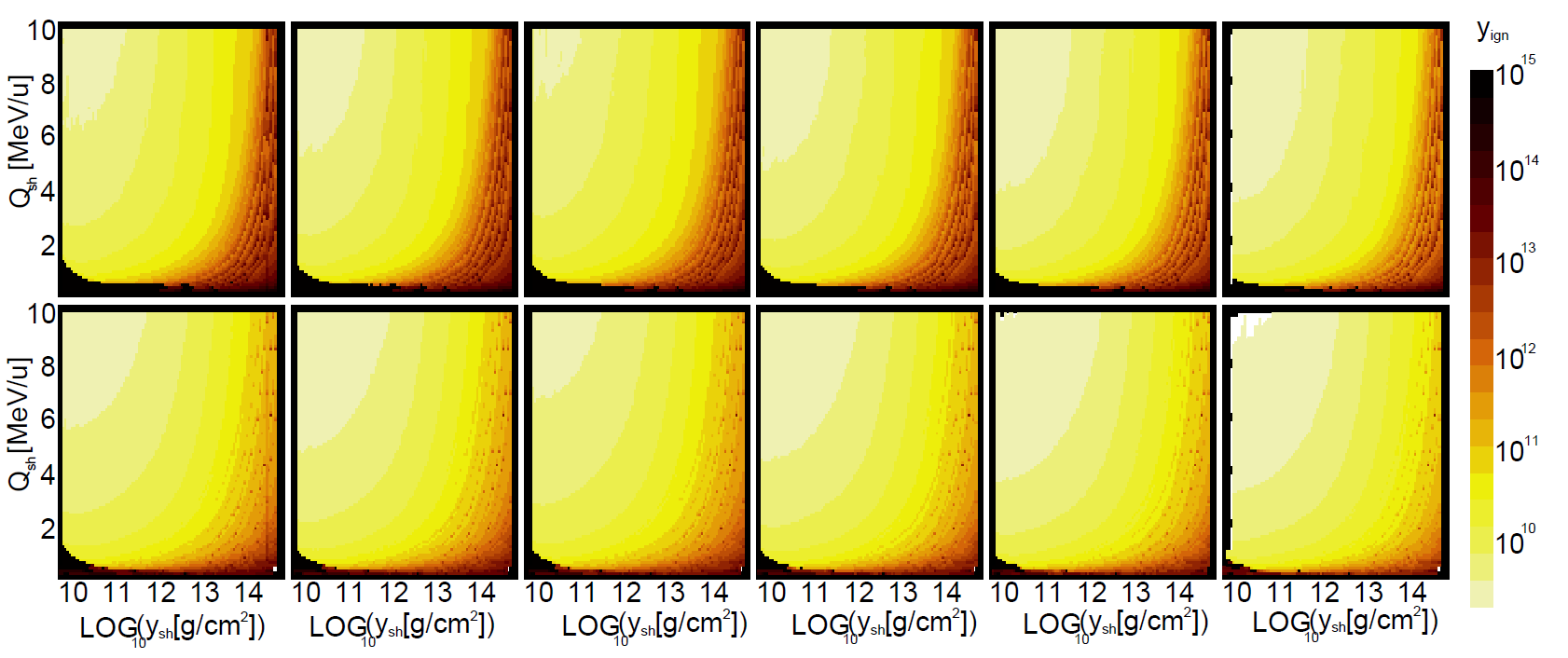}}
  \caption{Same as Figure~\ref{fig:yignmap}, highlighting the $Q_{\rm imp}$, $\dot{M}$ case where urca cooling has the most significant impact on the $y_{\rm ign}$ landscape.}
 \label{fig:yignmapurca}
\end{figure*}

\section{Results}
\label{sec:results}

Figure~\ref{fig:yignmap} shows $y_{\rm ign}$ resulting from Section~\ref{ssec:ignitiondepth} for calculations performed with the no urca cooling scenario. The columns of the subfigures show the impact of adopting different $^{12}{\rm C}+^{12}$C reaction rates, while the rows of the subfigures show the impact of $\Delta t$. The solid-black regions in these figures correspond to cases where the conditions for $y_{\rm ign}$ were not met, i.e. the carbon ignition curve and thermal profile did not intersect within $y=10^{10}-10^{15}$~g\,cm$^{-2}$. For all cases, $y_{\rm ign}$ is relatively shallow for large $Q_{\rm sh}$ and shallow $y_{\rm sh}$, while $y_{\rm ign}$ is relatively deep for the opposite scenario. Contours of approximately equal $y_{\rm ign}$ follow a trajectory of increasing $Q_{\rm sh}$ and deepening $y_{\rm sh}$. However, the contours abruptly end for low $Q_{\rm sh}$ and shallow $y_{\rm sh}$. This is because insufficient heat is deposited and retained within the neutron star outer layers in order for the thermal profile to intersect the ignition curve. For low $\dot{M}$, this region spans a considerable portion of the phase-space, even failing to achieve carbon ignition for $Q_{\rm sh}\approx9$~MeV\,$u^{-1}$ if $y_{\rm sh}\approx10^{10}$~g\,cm$^{-2}$. 

The gradient of $y_{\rm ign}$ within the $Q_{\rm sh}$-$y_{\rm sh}$ phase space becomes steeper moving from large $Q_{\rm sh}$ and shallow $y_{\rm sh}$ to small $Q_{\rm sh}$ and deep $y_{\rm sh}$. This can be understood by considering Figure~\ref{fig:ExampleProfiles}. The thermal profile for small $Q_{\rm sh}$ has a relatively shallow slope $\partial T/\partial y$, while $\partial T/\partial y$ rapidly increases in magnitude for increasing $Q_{\rm sh}$, approaching a converged slope. For deep $y_{\rm sh}$, the thermal profile intersects the carbon ignition curve at deep $y$, where the carbon ignition curve slope is especially shallow. As such, for small $Q_{\rm sh}$ and deep $y_{\rm sh}$, $y_{\rm ign}$ depends on the the intersection of two shallow-sloped curves, which will be particularly sensitive to small changes in the slope of the thermal profile, leading to a more rapid change in $y_{\rm ign}$ in this region of the $Q_{\rm sh}$-$y_{\rm sh}$ phase-space. 

\begin{figure*}
	\includegraphics[width=1.35\columnwidth]{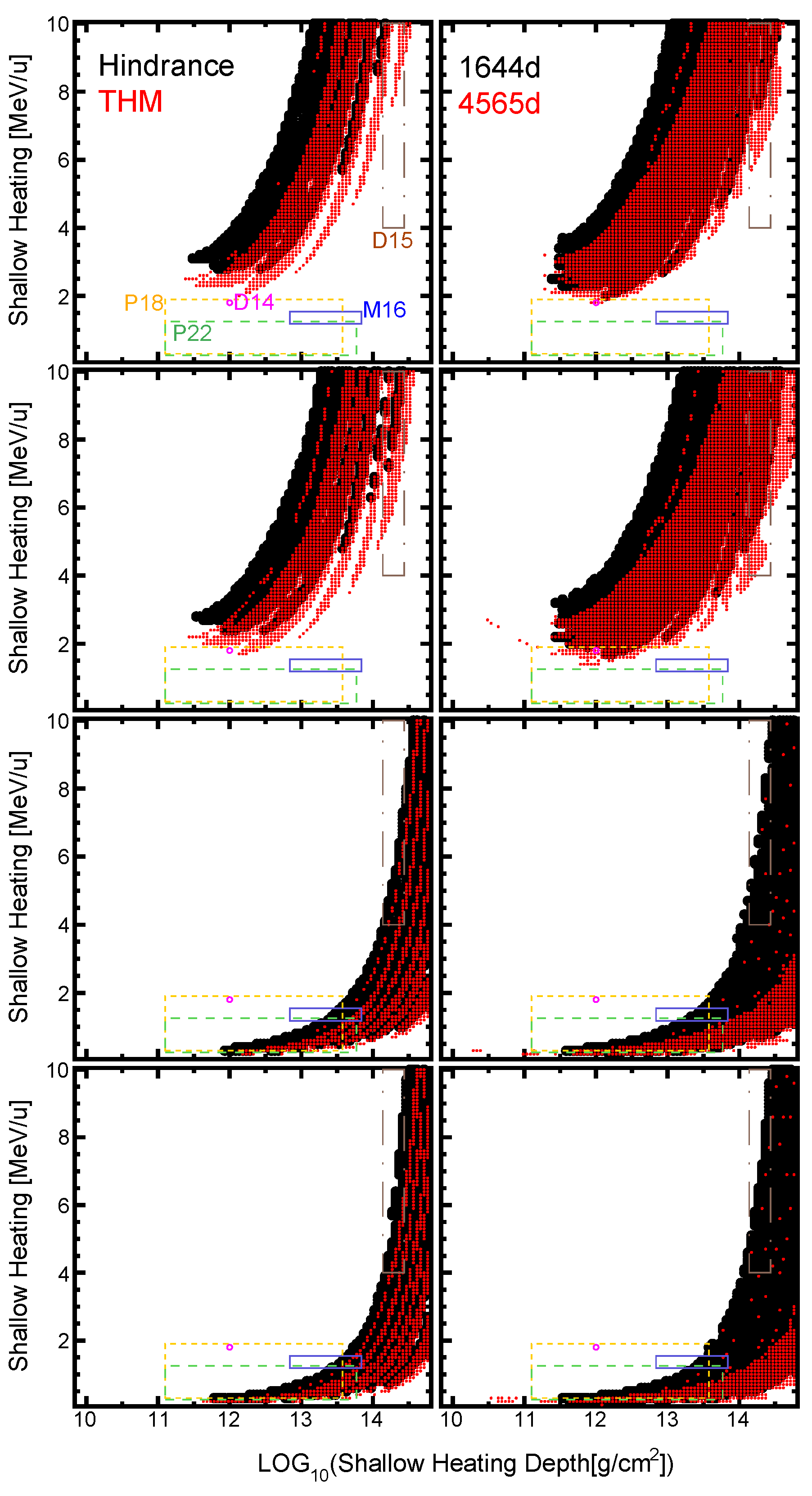}
    \caption{Map of regions in which the calculated $y_{\rm ign}$ is within the inferred range $y_{\rm ign,obs}=0.5-3\times10^{12}$~g\,cm$^{-2}$, where rows correspond to the same model conditions as the rows of Figure~\ref{fig:yignmap}. In the left column, the solid-black and dotted-red regions correspond to using the Hindrance and THM $^{12}{\rm C}+^{12}$C rates, respectively, with $\Delta t=1643.6$~d. In the right column, solid-black and dotted-red regions correspond to $\Delta t=1643.6$~d and 4565~d, respectively, where results for all $^{12}{\rm C}+^{12}$C rates are shown. Constraints on $Q_{\rm sh}$ and $y_{\rm sh}$ obtained from fits to crust cooling sources are shown for comparsion, where EXO 0748-676 (D14~\citep{Dege14}) is the pink open circle, MXB 1659-29 (P18~\citep{Pari18}) is the yellow short-dashed box, KS 1731-26 (M16~\citep{Merr16}) is the blue solid box, and MAXI J0556-332 (D15~\citep{Deib15}, P22~\citep{Page22}) constraints are the brown dot-dash and green long-dashed boxes, respectively.}
    \label{fig:yignconstraints}
\end{figure*}

The results shown in Figure~\ref{fig:yignmap} demonstrate a weak sensitivity to $Q_{\rm imp}$ and $^{12}{\rm C}+^{12}$C rate, as well as a modest sensitivity to $\Delta t$. Of all parameters investigated in this work, $y_{\rm ign}$ is primarily sensitive to $\dot{M}$. While both $\dot{M}$ and $\Delta t$ impact $Q_{\rm heat}$ deposited within the crust, $\Delta t$ controls how close the thermal profile is to steady-state, while $\dot{M}$ decides what the steady-state thermal profile is. Here, both $\Delta t$ were sufficiently close to achieving steady-state that the difference in thermal profiles is relatively modest. This would not necessarily be the case for an abnormally short $\Delta t$, as observed for 4U 1608-522~\citep{Keek08}. Similarly, $Q_{\rm imp}$ impacts $\kappa$ and therefore the thermal diffusion time, but does not have a particularly strong impact on the steady-state thermal profile.

Results were very nearly identical when including urca cooling. The only scenario in which urca cooling was observed to have some impact was $Q_{\rm imp}=40$ $\dot{M}=1.75\times10^{-8}$~$M_{\odot}$\,yr$^{-1}$. This (extremely modest) impact is shown in Figure~\ref{fig:yignmapurca}, where implementing the maximum urca cooling scenario slightly changes $y_{\rm ign}$ at low $Q_{\rm sh}$ and deep $y_{\rm sh}$. Upon close inspection, one sees urca cooling drives $y_{\rm ign}$ slightly deeper around $Q_{\rm sh}\approx2$~MeV\,$u^{-1}$, $\log(y_{\rm sh}[{\rm g\,cm}^{-2}])\approx14.5$ for $\Delta t=4565$~d. Consulting Figure~\ref{fig:ExampleProfiles}, it is apparent that the impact of shallow heating on the thermal profile is mostly concentrated near $y_{\rm sh}$. Given the strong $T$-dependence of $L_{\nu}$ (see Equation~\ref{eqn:Lnu}), urca cooling will result in an insignificant $Q_{\rm cool}$ unless $y_{\rm sh}$ approaches $y_{\rm urca}$, consistent with the findings of~\citet{Deib16}. Note that the present work only considers the impact of the $^{55}{\rm Sc}-^{55}$Ca urca pair, which is located in the crust. The impact of urca cooling may be more significant when considering an urca pair located in the accreted neutron star ocean, though there the strong $Q_{\rm EC}$-dependence will reduce $L_{\nu}$ (see Equation~\ref{eqn:L34}) and therefore $Q_{\rm cool}$.

Figure~\ref{fig:yignconstraints} shows the regions from Figure~\ref{fig:yignmap} where $y_{\rm ign}$ calculated in this work is within the ignition depth inferred from model-observation comparisons of superburst light-curves, $y_{\rm ign,obs}=0.5-3\times10^{12}$~g\,cm$^{-2}$~\citep{Cumm06}. The left and right columns of this figure show the influence of the $^{12}{\rm C}+^{12}$C reaction rate and $\Delta t$, respectively. For comparison, Figure~\ref{fig:yignconstraints} includes constraints on shallow heating obtained from fits to crust cooling light curves performed in earlier works. These include the results of fits to EXO 0748-676~\citep{Dege14}, MXB 1659-29~\citep{Pari18}, KS 1731-26~\citep{Merr16}, and MAXI J0556-332~\citep{Deib15,Page22}, of which it is noted that only KS 1731-26 has been observed to feature superbursts. The two constraints for MAXI J0556-332 are quite disparate, as \citet{Page22} assumed that the first crust cooling event observed for this source was shortly preceded by a hyperburst within the crust, while \citet{Deib15} fit only this cooling event while assuming shallow heating was responsible for the large energy deposition in the neutron star outer layers. For the low $\dot{M}$ modeled in this work, which is closer to the $\dot{M}$ observed prior to superbursts~\citep{intZ17}, the shallow heating constraints obtained in the present work are on the edge of consistency with crust cooling constraints. Here, for the lower of the two $\dot{M}$, the present work favors generally larger $Q_{\rm sh}$ and deeper $y_{\rm sh}$, with the exception of the unique findings of \citet{Deib15}. For $\dot{M}$ near the Eddington limit, the present work results in shallow heating that is in agreement with most constraints from crust cooling model-observation comparisons, but skewing to deeper $y_{\rm sh}$. For large $\dot{M}$, there is stronger sensitivity of $y_{\rm ign}$ to $\Delta t$ at deep $y_{\rm sh}$.

\section{Discussion}
\label{sec:discussion}

\subsection{Shallow Heating Constraints}
\label{ssec:shallowheatconstraints}

Figure~\ref{fig:yignconstraints} demonstrates that the inferred depth of carbon ignition for superbursts can be used as a constraint for the magnitude and depth of shallow heating in accreting neutron stars. This constraint is primarily sensitive to $\dot{M}$, modestly sensitive to $\Delta t$ and the $^{12}{\rm C}+^{12}$C reaction rate, weakly dependent on $Q_{\rm imp}$, and negligibly dependent on urca cooling in the accreted crust. The primary sensitivity, $\dot{M}$, can usually be constrained for a superbursting system based on the persistent luminosity (though there are complications related to X-ray reflection from the accretion disk)~\citep{He16}. Therefore, this shallow heating constraint is relatively robust to changes in modeling assumptions.

Relative to the complementary constraints on shallow heating from crust cooling model-observation comparisons, the superburst ignition depth constraint is weaker in that it allows for a larger region of the $Q_{\rm sh}$-$y_{\rm sh}$ phase-space. This is to be expected, as $y_{\rm ign}$ essentially only depends on the thermal structure at a single $y$, while crust cooling light curves depend on the thermal structure of the entire neutron star. However, the superburst ignition depth constraint offers the distinct advantage of being insensitive to $Q_{\rm imp}$ and urca cooling, contrary to crust cooling light curves~\citep{Brow09,Meis17}. Furthermore, most superbursting sources do not feature crust cooling episodes~\citep{Gall20} and therefore the method presented here can expand the set of accreting neutron stars that can be used to constrain shallow heating.

Of the crust-cooling sources featured in Figure~\ref{fig:yignconstraints}, only KS 1731-26 has also exhibited a superburst. The $\dot{M}$ inferred from the persistent X-ray luminosity prior the the superburst event is $\approx10$\% of Eddington~\citep{Kuul02}, corresponding to the lower $\dot{M}$ modeled in the present work. This $\dot{M}$ is in agreement with the $\dot{M}$ that one would infer based on the recurrence time between KS1731-26 X-ray bursts prior to the superburst~\citep{Lamp16,Meis19}. Fits to the KS 1731-26 crust-cooling light curve infer that $Q_{\rm imp}\approx4$~\citep{Merr16,Lali19}. The $\Delta t$ prior to the quiescent episode of KS 1731-26 in 2001 is assumed to be 4565~d based on observational data~\citep{Merr16}, implying that $\Delta t$ prior to the 1998 superburst event was $\Delta t\approx$2900~d. Therefore, the results from the present work that are most relevant for KS 1731-26 are from the calculations with $Q_{\rm imp}=4$ and $\dot{M}=1.75\times10^{-9}$~$M_{\odot}$\,yr$^{-1}$. These results do not appear to be consistent with the shallow-heating constraints obtained by crust-cooling light curve fits in \citet{Merr16} (blue solid box in Figure~\ref{fig:yignconstraints}). However, given the approximate nature of the $\dot{M}$ constraints prior to the suprbursting episode, it is plausible that consistency could be achieved. More detailed analysis, e.g. of the persistent X-ray luminosity prior to the superburst or model-observation comparisons for the light curve shape and recurrence time of standard bursts~\citep{Meis18b,John20}, is likely necessary. Additionally, while the crust-cooling models of \citet{Merr16} fit for $Q_{\rm sh}$, they held $y_{\rm sh}$ fixed. It is therefore possible that the inconsistency between their $Q_{\rm sh}$ constraints and this work are due to fixing $y_{\rm sh}$ in that work. Further crust-cooling model-observation comparisons are needed to draw stronger conclusions.

\subsection{Influence of Nuclear Physics Uncertainties}
\label{ssec:nucuncertainties}

The shallow heating constraints obtained in this work are relatively insensitive to assumptions regarding input nuclear physics. This includes not only the $^{12}{\rm C}+^{12}$C nuclear reaction rate, but also past surface burning and nuclear reactions occurring within the accreted crust.

As shown in Figures~\ref{fig:ExampleProfiles} and \ref{fig:yignmap}, adopting different $^{12}{\rm C}+^{12}$C reaction rates leads to minor changes in the calculated $y_{\rm ign}$. When comparing the two most discrepant theoretical predictions for this rate, which differ in $S^{*}$ by nearly six orders of magnitude~\citep{Tang22}, the change in $y_{\rm ign}$ is roughly a factor of four. This relative insensitivity is due to the extreme temperature dependence of the $^{12}{\rm C}+^{12}$C rate, owing to the considerable Coulomb barrier. Nonetheless, the impact is on the same order of the uncertainty in the inferred $y_{\rm ign,obs}$ and so some uncertainty reduction in the $^{12}{\rm C}+^{12}$C rate would be beneficial. It would be particularly beneficial to push direct measurements of this nuclear reaction cross section down to slightly lower energies in order to confirm or exclude the THM~\citep{Tumi18} and Hindrance~\citep{Jian18} rates, which are responsible for the bulk of the impact found in the present work. Based on the relative $S^{*}$ of theoretical models, it appears that direct measurements of $^{12}{\rm C}+^{12}$C down to $E=2.25$~MeV may suffice. Though~\citet{Tan20} reach $E=2.20$~MeV for their lowest-energy measurement, that data-point is a relatively large upper-limit on $S^{*}$ and has a large $E$-separation from the neighboring points measured in that work, making it difficult to confront with theoretical predictions.

For all accreting neutron star systems, the crust composition is uncertain due to uncertainties in which surface-burning modes (i.e. stable burning, superbursts, or X-ray bursts) were prevalent in the past, the nuclear physics of the surface-burning processes, and nuclear physics of the accreted crust. For instance, superburst ashes and X-ray burst ashes imply a substantially different $Q_{\rm imp}$~\citep{Meis18}. Meanwhile, individual nuclear reaction rates, and even individual nuclear masses, can have important impacts on $Q_{\rm imp}$ and $X_{i}$ of urca nuclides~\citep{Cybu16,Scha17,Ong18,Meis19,Hoff20,Meis22}. Nuclear reactions in the crust further modify the crust composition, but these modifications depend sensitively on input nuclear physics, such as nuclear masses and the presence of exotic reaction processes~\citep{Shch19,Scha22b}. Furthermore, even if $X_{i}$ were known, the $L_{\nu}$ themselves are sensitive to the adopted $Q_{\rm EC}$ and $ft$, which are often not known~\citep{Meis15b,Ong20}. The insensitivity of results in the present work regarding the adopted $Q_{\rm imp}$ and urca cooling imply that these considerable uncertainties in the crust composition and $L_{\nu}$ are mostly inconsequential. The caveat to this statement is that the present work only considered an urca pair in the crust, which may not apply to an urca pair located closer to $y_{\rm sh}$ in the ocean. Furthermore, it is important to highlight that the surface-burning production of $^{12}$C is extremely important, as it sets $X_{\rm C}$ in the ignition curve calculations. The mechanism to produce sufficiently high $X_{\rm C}$ is uncertain, but appears to require the system to spend some time in a special region of $\dot{M}$~\citep{Stev14,Keek16}.

The remaining nuclear physics uncertainties that were not investigated in this work include $e^{-}$-capture heating and deep crustal heating, where the latter is primarily dependent on the crust-core transition pressure and hence the dense-matter equation-of-state~\citep{Shch21,Shch22}. Neither of these heat sources would remove the need for shallow heating, but they do impact the accreting neutron star thermal profile and therefore likely have some influence on the shallow heating constraints inferred from $y_{\rm ign}$~\citep{Coop09}.

Other important microphysics uncertainties that were not investigated in the present work include the Coulomb logarithm and plasma-screening for carbon ignition, as well as the direct urca process for the crust thermal profile. Neither the Coulomb logarithm nor plasma screening effects are accessible in terrestrial laboratory experiments in the foreseeable future (with the exception of plasma-screening effects for nuclear reactions of light nuclides at modest densities~\citep{Kemp19}) and therefore dedicated theoretical efforts will be required. Once the physical origin of shallow heating is eventually determined, it is possible that $y_{\rm ign}$ model-observation comparisons like the ones presented here can provide some constraints. The impact of the direct urca process on superburst ignition was recently investigated by \cite{Dohi22}. Those authors found that direct urca operating in the neutron star core would increase the carbon ignition depth, possibly suppressing superbursts altogether, depending on the astropysical conditions.

\subsection{Incorporation into Multi-observable Modeling}
\label{ssec:multiobs}

The method for obtaining shallow heating constraints presented in this work could be applied to any accreting neutron star system featuring superbursts. However, the value of these constraints would be increased by combining them with constraints derived from other observables. The  source KS 1731-26 is particularly promising in this regard, as it features X-ray bursts that resemble the successfully modeled~\citep{Meis18,John20} bursts of GS 1826-24~\citep{Muno00}, photospheric radius expansion bursts that can be used to get $M_{\rm NS}$ and $R_{\rm NS}$ constraints~\citep{Ozel12}, superbursts~\citep{Kuul02}, and an episode of crust cooling~\citep{Rutl02}.

Ideally, multi-observable modeling of a source such as KS 1731-26 would use consistent assumptions for the system properties across the models. However, as shown in this work, the exact crust composition need not be used for shallow heating constraints from calculations of $y_{\rm ign}$. Ignoring this complication would reduce the computational cost of incorporating the $y_{\rm ign}$ shallow heating constraints into the multi-observable modeling.

\section{Conclusions}
\label{sec:conclusions}

The present work demonstrates that the inferred depth of carbon ignition for X-ray superbursts can be used to constrain the depth and magnitude of shallow heating in the accreted neutron star crust. This constraint is shown to be weakly sensitive to input nuclear physics, including assumptions regarding the $^{12}$C+$^{12}$C nuclear reaction rate and the crust composition. The main model sensitivity is the accretion rate $\dot{M}$ prior to the superburst, along with the accretion outburst duration $\Delta t$, if $\Delta t$ is comparable to thermal time at the depth of the shallow heat source. This method provides a new way to constrain shallow heating, expanding the number of accreting neutron star sources that can be used for this purpose. For sources featuring other observables such as crust cooling, the inclusion of this method into multi-observable modeling may improve the stringency of constraints on shallow heating.

\section*{Acknowledgements}
The views expressed in this article are those of the author and do not reflect the official guidance or position of the United States Government, the Department of Defense, the United States Air Force, or the United States Space Force.
I thank Hendrik Schatz, Wei Jia Ong, and Duncan Galloway for useful discussions, Xiadong Tang for providing tables of the $^{12}$C+$^{12}$C $S^{*}$-factors, Ed Brown for creating and maintaining a public release of {\tt dStar}, and the {\tt MESA} developers for creating and maintaining a public release of that code and the associated software development kit.
 This work was inspired by conversations at a workshop that was supported by funds from the U.S. National Science Foundation under Grants No. PHY-1430152 (Joint Institute for Nuclear Astrophysics -- Center for the Evolution of the Elements) and OISE-1927130 (International Research Network for Nuclear Astrophysics).

\section*{Data Availability}
The Appendix contains a sample {\tt dStar} inlist that could be used to recreate the thermal profiles used in this work. The carbon ignition curves are available as Supplementary Material.



\bibliographystyle{mnras}
\bibliography{references} 




\appendix

\section{Sample {{\tt dStar}} Inlist}
\label{sec:appendixA}

The following is an example inlist for {\tt dStar}~\citep{Brow15} calculations of the accreted crust thermal profile performed for the present work.

\begin{lstlisting}[breaklines=true,basicstyle=\tiny]
&controls
     
    ! controls for output 
    write_interval_for_terminal = 10000 
    write_interval_for_terminal_header = 10000 
    write_interval_for_history = 1 
    write_interval_for_profile = 1 
    starting_number_for_profile = 1 
     
    output_directory = 'LOGS' 
     
    ! controls for the solver 
    maximum_number_of_models = 10000 
    maximum_timestep = 0.0 
        ! implies that max = tend-t 
    integration_tolerance = 1.0d-4 
    ! limits on temperature: if a zone goes outside these bounds, reduce stepsize 
    min_lg_temperature_integration = 7.0 
    max_lg_temperature_integration = 9.5 
    ! spatial resolution 
    target_resolution_lnP = 0.05 
     
    ! macroscopic NS parameters 
    fix_core_temperature = .TRUE. 
    core_temperature = 1.e8 
    fix_atmosphere_temperature_when_accreting = .FALSE. 
    atmosphere_temperature_when_accreting = 2.4d8 
     
    ! integration epochs 
    number_epochs = 4 
    ! 4 epochs, time in days. 
    !For 1.75e-8 Msol/yr eddington, 1.103e17g/s is 10pct Edd. 1.103e18g/s is 100pct Edd 
    basic_epoch_Mdots = 4*1.103e+17
    !Time to accrete to y=12g/cm2 for EOS radius below at 100%Edd,10%Edd; then KS1731 outburst time 
    basic_epoch_boundaries = 0.0,164.4,1643.6,4565.0 
     
    ! core properties for 12.31km 1.4msol NS using APR EOS, as in Lalit+2016 
    core_mass = 1.376     ! Msun 
    core_radius = 11.227    ! km 
    ! crust boundaries (pressure) 
    lgPcrust_bot = 32.819 ! cgs 
    lgPcrust_top = 22.0 ! cgs 
     
     
    which_neutron_1S0_gap = 'sfb03' 
     
    ! atmosphere composition 
    lg_atm_light_element_column = 4.0 
     
    ! impurities 
    fix_Qimp = .TRUE. 
    Qimp = 4
     
    ! heating 
    turn_on_extra_heating = .TRUE. 
    Q_heating_shallow = 9.9
    lgP_min_heating_shallow = 28.47
    lgP_max_heating_shallow = 29.43
     
    ! shell Urca cooling 
    turn_on_shell_Urca = .TRUE. 
    shell_Urca_luminosity_coeff1 = 4.8
    lgP_shell_Urca1 = 29.159
     
/
\end{lstlisting}


\bsp	
\label{lastpage}
\end{document}